\documentclass[a4paper,twocolumn,showpacs,nofootinbib,floatfix,prc]{revtex4}

\usepackage{graphicx}
\usepackage{amsfonts}
\renewcommand{\vec}[1]{\mathbf{#1}}
\def\be{\begin{equation}}
\def\ee{\end{equation}}
\def\bea{\begin{eqnarray}}
\def\eea{\end{eqnarray}}
\newcommand{\beq}{\begin{equation}}
\newcommand{\eeq}[1]{\label{#1} \end{equation}}

\begin{document}
%%%%%%%%%%%%%%%%%%%%%%%%%%%%%%%%%%%%%%%%%%%%%%%%%%%%%%%%%%%

\title{Kinetic description of particle emission from expanding source}

\date{29.01.2008}

\author{V.K. Magas$^1$, L.P. Csernai$^{2,3}$}

\affiliation{
$^1$ Departament d'Estructura i Constituents de la Mat\'eria,\\
Universitat de Barcelona, Diagonal 647, 08028 Barcelona, Spain\\
$^2$ Theoretical and Energy Physics Unit,\\ 
University of Bergen, Allegaten 55, 5007 Bergen, Norway \\
$^3$ MTA-KFKI, Research Inst of Particle and Nuclear Physics,\\
H-1525 Budapest 114, P.O.Box 49, Hungary
}

\begin{abstract}
{
The freeze out of the expanding systems, created in relativistic heavy ion collisions, is discussed. We combine kinetic  freeze out equations 
with Bjorken type system expansion into a unified model. The important feature of the proposed scenario is that physical freeze out is 
completely finished in a finite time, which can be varied from $0$ (freeze out hypersurface) to 
$\infty$. The dependence of the post freeze out distribution function on the freeze out time will be studied. As an example, model is completely solved and analyzed for the gas of
pions. We shall see that the basic freeze out features, pointed out in the earlier works, are not smeared out by the expansion of the system. The entropy evolution in such a scenario is also studied. 

}
\end{abstract}


\maketitle

%%%%%%%%%%%%%%%%%%%%%%%%%%%%%%%%%%%%%%%%%%%%%%%%%%%%%%%%%%%%%%%%%%%%%%%%%%%%%%%%%%%%%%%%%%%
\section{Introduction}
%%%%%%%%%%%%%%%%%%%%%%%%%%%%%%%%%%%%%%%%%%%%%%%%%%%%%%%%%%%%%%%%%%%%%%%%%%%%%%%%%%%%%%%%%%%

In the ultrarelativistic heavy ion collisions at RHIC the total number of the produced particles exceeds several thousands, %\cite{mult}, 
therefore one can expect that the produced system behaves as a "matter" and generates collective effects. Indeed strong collective flow patterns have been measured at RHIC, %\cite{flow}, 
which suggests that  
the hydrodynamical models are well justified during the intermediate stages of the reaction: from the time when local equilibrium is reached until the freeze out (FO),
when the hydrodynamical description breaks down.
During this FO stage, the matter becomes so dilute and cold that  particles stop interacting and
stream towards the detectors freely, their momentum distribution freezes out.
The FO stage is essentially the last part of a collision process and the main source for observables.

In simulations FO is usually described in two extreme ways: either FO on a hypersurface with zero thickness, or FO described by volume emission model or hadron cascade, which  require an infinite time and space for a complete FO. At first glance it seems that one can avoid troubles with FO modeling using hydro+cascade two module model \cite{hydro_cascade}, since in hadron cascades gradual FO is realized automatically. However, in such a scenario there is an uncertain point, actually uncertain hypersurface, where one switches from hydrodynamical to kinetic modeling. First of all it is not clear how to determine such a hypersurface. This hypersurface in general may have both time-like and space-like parts. Mathematically this problem is very similar to hydro to post-FO transition on the infinitely narrow FO hypersurface, therefore all the problems discussed for FO on the hypersurface with space-like normal vectors will take place here. Another complication is that while for the post FO domain we have mixture of non-interacting ideal gases, now for the hadron cascade we should generate distributions for the interacting hadronic gas of all possible species, as a starting point for the further cascade evolution. The volume emission models are also based on the kinetic equations \cite{vol_em,grassi04,vol_em_2,old_SL_FO_2} 
defining the evolution of the distribution functions, and therefore these also require to generate initial distribution functions for the interacting hadronic species on some hypersurface. 

In the recent works \cite{ModifiedBTE} it has been shown that the basic assumptions of the Boltzman Transport Equation (BTE) are not satisfied during FO and therefore its description has to be based on Modified BTE,  suggested in \cite{ModifiedBTE}.
When the characteristic length scale, describing the change of the distribution function, becomes smaller than mean free path (this always happens at late stages of the FO), then the basic
assumptions of the BTE get violated, and the expression for the collision integral has
to be modified to follow the trajectories of colliding particles, what makes calculations much more complicated.  
In fact, in cascade model one
follows the trajectories of the colliding particles, therefore what is effectively solved is not BTE, but Modified BTE.
Once the necessity of the Modified BTE and at the same time the difficulty of its direct solution were realized, the simplified kinetic FO models become even more important for the understanding of the principal features of this phenomenon. Such simplified models were developed and studied in Refs. \cite{old_SL_FO_2,old_SL_FO,old_TL_FO,Mo05a,Mo05b}, but all these were missing expansion. Thus, the important question to be studied is whether the important freeze out features, pointed out in the earlier works, are not smeared out by the expansion of the system?  

In this paper we present a simple kinetic FO model, which describes the freeze out of particles from a Bjorken expanding fireball \cite{Bjorken}.
Thus, this is a more physical extension of the oversimplified FO models without expansion \cite{old_SL_FO_2,old_SL_FO,old_TL_FO,Mo05a,Mo05b}. 
Taking the basic ingredients of the FO simulations from Refs. \cite{Mo05a,Mo05b} we can make physical freeze out to be  
completely finished in a finite time, which can be varied from $0$ (freeze out hypersurface) to $\infty$. In the other words our freeze out happens in a layer, i.e. in a domain restricted by two parallel 
hypersurfaces $\tau=\tau_1$ and $\tau=\tau_1+L$ ($\tau$ is the proper time).

In Ref. \cite{old_TL_FO} authors have also adopted kinetic gradual FO model to Bjorken geometry, but combined it with Bjorken expansion on the consequent, not on the parallel basis: system expands according to Bjorken hydro scenario, but when it reaches beginning of the FO process system stops expansion and gradually freezes out in a fixed volume. However, it was shown in \cite{old_TL_FO} that such a scenario is not physical: the simultaneous modeling of expansion and freeze out is required in order to avoid decreasing  total entropy. Now we propose such a generalized model.

%%%%%%%%%%%%%%%%%%%%%%%%%%%%%%%%%%%%%%%%%%%%%%%%%%%%%%%%%%%%%%%%%%%%%%%%%%%%%%
\section{Finite layer freeze out description}
%%%%%%%%%%%%%%%%%%%%%%%%%%%%%%%%%%%%%%%%%%%%%%%%%%%%%%%%%%%%%%%%%%%%%%%%%%%%%%

Let us briefly review gradual FO model, which we are going to generalize including expansion. Many building blocks of the model are Lorentz invariant and can be applied to both time-like and space-like FO layers, so at the beginning we will write these in general way. 
Starting from the Boltzmann Transport Equation, introducing two components of the distribution function, $f$: the interacting, $f^i$, and the frozen out,
$f^f$  ones, ($f=f^i+f^f$), and assuming that FO is a directed process (i.e. neglecting the gradients of the distribution functions in the directions perpendicular to the FO direction comparing to that in the FO direction) we can obtain the following system of the equations \cite{ModifiedBTE,Mo05a}: 
\be
\frac{d f^{i}}{d s}  = - \frac{P_{esc} }{\tau_{FO}}f^{i} + \frac{ f_{eq}(s) - f^{i} }{\tau_{th}} \, ,
\quad
\frac{d f^{f}}{d s}  =  \frac{P_{esc}}{\tau_{FO}} f^{i} \, .
\label{sys}
\ee
The FO direction is defined by the unit vector $d\sigma_\mu$. FO happens in a layer of given thickness $L$ with two parallel boundary hypersurfaces perpendicular to $d\sigma_\mu$, and $s=d\sigma_\mu x^\mu$ is a length component in the FO direction. We work in the reference frame of the front, where $d\sigma_\mu$ is either $(1,0,0,0)$ for the time-like FO, or $(0,1,0,0)$ for the space-like FO. The $\tau_{FO}$ is some characteristic
length scale, like mean free path or mean collision time for time-like FO. The rethermalization of the interacting component is taken into account via the relaxation time approximation,
where $f_i$ approaches the equilibrated J\"uttner distribution, $f_{eq}(s)$, with a
relaxation length/time, $\tau_{th}$. The system (\ref{sys}) can be solved semi-analytically
in the fast rethermalization limit \cite{Mo05a}.

The basis of the model, i.e. the
invariant escape rate of the particles within the FO layer of the thickness $L$, 
for both time-like and space-like normal vectors is given as (see Refs. \cite{Mo05a,Mo05b,kemer} for more details)
\be
   P_{esc} =
   \left( \frac{L}{L-s} \right)
   \left(\frac {p^\mu d\sigma_\mu}{p^\mu u_\mu}\right)\
   \Theta(p^\mu d\sigma_\mu)\,,
\label{esc1}
\ee
where $p^\mu$ is the particle four-momentum, $u^\mu$ is the flow velocity. 
In fact the model based on the escape rate (\ref{esc1}) is a generalization of a simple kinetic models studied in Refs. \cite{old_SL_FO_2,old_SL_FO,old_TL_FO}, which
can be restored in the $L\rightarrow \infty$ limit.
Here we will concentrate on the time-like case only, where the above $\Theta$ function
is unity. 

The important feature of the escape rate in the form  (\ref{esc1}) is that physical freeze out is 
completely finished when $s=L$, i.e. it requires finite space/time extent. Furthermore, now we can vary this layer thickness, $L$, from $0$ (freeze out hypersurface) to 
$\infty$ and study  how the post FO distribution depends on the layer thickness. Interesting and unexpected result was found in \cite{Mo05a,Mo05b,kemer}, for both space-like and time-like FO layers, namely that if $L$ is large enough, at least several $\tau_{FO}$, then post FO distribution gets some universal form, independent on the layer thickness.

Simple semianalytically solvable FO models studied in \cite{old_SL_FO_2,old_SL_FO,old_TL_FO,Mo05a,Mo05b} are missing an important ingredient - the expansion of the freezing out system. The open question is whether the features of the FO discussed above will survive if the system expansion is included. In this work we are going to build a model, which includes both gradual FO and Bjorken-like expansion of the system, and answer this question. 

%%%%%%%%%%%%%%%%%%%%%%%%%%%%%%%%%%%%%%%%%%%%%%%%%%%%%%%%%%%%%%%%%%%%%%%%%%%%%%
\section{Bjorken expansion with gradual freeze out}
%%%%%%%%%%%%%%%%%%%%%%%%%%%%%%%%%%%%%%%%%%%%%%%%%%%%%%%%%%%%%%%%%%%%%%%%%%%%%%

Let us first remind the reader the basics of the Bjorken model \cite{Bjorken}. Bjorken model is one-dimensional in the same sense as discussed before eq. (\ref{sys}) - only the proper time, $\tau=\sqrt{t^2-x^2}$, gradients are considered. Here the reference frame of the front, where $d\sigma^\mu=(1,0,0,0)$, is the same as the local rest frame (comoving frame), where $u^\mu=(1,0,0,0)$. The evolution of the energy density and baryon density is given by the following equations:
\beq
\frac{d e}{d \tau}=-\frac{e+P}{\tau}\,, \quad \frac{d n}{d \tau}=-\frac{n}{\tau}\,,
\eeq{Bjorken}   
where $P$ is the pressure. 
The initial conditions are given at some $\tau=\tau_0$: $e(\tau_0)=e_0$, $n(\tau_0)=n_0$. This system can be easily solved:
\beq
e(\tau)=e_0\left(\frac{\tau_0}{\tau}\right)^{1+c_o^2}\,, \quad
n(\tau)=n_0\left(\frac{\tau_0}{\tau}\right)\,,
\eeq{pure_bjor}
where $P=c_o^2 e$ is the equation of state (EoS) with constant velocity of sound, $c_o$.

It is important to remember that if we want to have a finite volume fireball, we need to put some boarders on the system. Here we assume that our system, described by the 
Bjorken model, is situated in the spacial domain $|\eta|\le \eta_{R}$ or what is the same $|z|\le z_R(\tau) = \tau \sinh \eta_{R}$ ($\eta=\frac{1}{2}\ln \left(  \frac{t+z}{t-z}\right)$ is a space-time rapidity). Within these boarders system is uniform along $\tau = const$ hyperbolas due to model assumptions, while outside we have vacuum with zero energy and baryon densities as well as pressure. Thus, we have a jump, a discontinuity on the boarder, which stays there during all the evolution. Certainly, to prevent matter expansion through such a boarder (due to strong pressure gradient) some work is done on the boarder surface \cite{Bjorken_work}. One can think about it as putting some pressure to the surface with the vacuum, exactly the one which would remove discontinuity, then work is done by the expanding system against this pressure. 
As the system expends the volume of the fireball increases as
\beq
V(\tau)=2 A_{xy} \sinh \eta_{R}\tau \,,
\eeq{vol}
where $A_{xy}$ is the transverse area of the system. 
It has been shown in \cite{Bjorken_work} that for such a finite Bjorken system the total energy is given by 
\beq
E(\tau)= 2 A_{xy} \sinh \eta_{R} \tau e(\tau)
\eeq{Etot} 
while the total entropy and the conserved charge appear to be 
\beq
S(\tau)= 2 A_{xy} \eta_{R} \tau s(\tau) \,, \quad N(\tau)= 2 A_{xy} \eta_{R} \tau n(\tau)\,.
\eeq{Stot} 
Work done by the expanding system, $W$, is given by 
\beq
dW=PdV \quad \Rightarrow \quad W(\tau)=e_0 V(\tau_0) \left( 1-\left(\frac{\tau_0}{\tau}\right)^{c_o^2}\right)\,.
\eeq{W}
One can easily check then the energy conservation:
\beq
E(\tau)+W(\tau)=e_0 V(\tau_0) = E(\tau_0)=const\,.
\eeq{e_con_1}

Applying our FO model to such a system, we obtain:
\beq
df^i(\tau')=-\frac{d\tau'}{\tau_{FO}}\frac{L}{L-\tau'}f^i(\tau')+\frac{d\tau'}{\tau_{th}}
\left[f_{eq}(\tau')-f^i(\tau')\right]\,,
\eeq{dfi}
\beq
df^f(\tau')=+\frac{d\tau'}{\tau_{FO}}\frac{L}{L-\tau'}f^i(\tau')\,,
\eeq{dff}
where FO begins at $\tau=\tau_1$ and $\tau'=\tau-\tau_1$. Taking the fast rethermalization limit, similarly to what is done in \cite{old_TL_FO}, we can obtain simplified equations for $f^i$, which is a thermal distribution $f^i(\tau)=f_{eq}(\tau)$, and $f^f$, and consequently for $e^i, n^i$ and $e^f, n^f$:
\beq
\frac{d e^i}{d \tau'}=-\frac{e^i}{\tau_{FO}}\frac{L}{L-\tau'}\,, \quad 
\frac{d n^i}{d \tau'}=-\frac{n^i}{\tau_{FO}}\frac{L}{L-\tau'}\,,
\eeq{2int}   
\beq
\frac{d e^f}{d \tau'}=+\frac{e^i}{\tau_{FO}}\frac{L}{L-\tau'}\,, \quad 
\frac{d n^f}{d \tau'}=+\frac{n^i}{\tau_{FO}}\frac{L}{L-\tau'}\,.
\eeq{2free}   

Now the idea is to create a system of equations which would describe a fireball which simultaneously expands and freezes out. Let us put our two components ($e=e^i+e^f$) into the first equation of (\ref{Bjorken}) and do some simple algebra:
\begin{equation}
\frac{d e^i}{d\tau}+
\frac{d e^f}{d\tau}=
-\frac{e^i+P^i}{\tau}-\frac{e^f}{\tau}-
\frac{e^i}{\tau_{FO}}\frac{L}{L-\tau'}+\frac{e^i}{\tau_{FO}}\frac{L}{L-\tau'}\,,
\label{fif}
\end{equation}
where last two terms add up to zero; the free component, of course, has no pressure. So far our eq. (\ref{fif}) is completely identical to the first equation of (\ref{Bjorken}). The main assumption of our model is that our system evolves in such a way that eq. (\ref{fif}) is satisfied as a system of two separate equations for interacting and free components \cite{Bjorken_FO,Bjorken_FO_mass}:
\beq
\frac{d e^i}{d \tau}=-\frac{e^i+P^i}{\tau}-\frac{e^i}{\tau_{FO}}\frac{L}{L+\tau_1-\tau}\,, 
\eeq{eint}   
\beq
\frac{d e^f}{d \tau}=-\frac{e^f}{\tau}+\frac{e^i}{\tau_{FO}}\frac{L}{L+\tau_1-\tau}\,.
\eeq{efree}   
Similarly we can obtain equations for baryon density \cite{Bjorken_FO,Bjorken_FO_mass}:
\beq
\frac{d n^i}{d \tau}=-\frac{n^i}{\tau}-\frac{n^i}{\tau_{FO}}\frac{L}{L+\tau_1-\tau}\,, 
\eeq{nint}   
\beq
\frac{d n^f}{d \tau}=-\frac{n^f}{\tau}+\frac{n^i}{\tau_{FO}}\frac{L}{L+\tau_1-\tau}\,.
\eeq{nfree}
In all these eqs. (\ref{eint}-\ref{nfree}) on r.h.s. there are two terms: the first one is due to expansion and the second one is due to freeze out.  

As it was discussed above our volume is restricted in $\eta$ space by two sharp boarders, $|\eta|\le \eta_R=const$.  All our interacting matter is maintained within these limits. However, the frozen out particles, which do not interact with system anymore, will not necessarily know about such a restriction and can, in principle, move faster than the boarder and thus will leave the volume under consideration.   
The momenta of the frozen out particles are distributed according to the distribution function, $f^f$, which (as we shall see later) falls down exponentially with momentum, while only the particles strongly directed along the beam line can have high rapidity. So, from the particles emitted close to midrapidity only a very tiny fraction will have momentum rapidity higher than $\eta_R$, and thus will leave our system volume. Particles freezing out from the system elements with higher space-time rapidity will have better, but still small, chance to leave the volume of consideration, and only the frozen out particles from the volume elements close to the $|\eta|= \eta_R$ boarders will disappear from our modeling with probability up to $50\%$, because $f^f$ is spherically symmetric in the comoving frame. For the reasonable $\eta_R$ boarders the amount of such particles can be neglected.
It is, however, clear that our approach is best justified for the reaction region near midrapidity. In this work we will do only illustrative calculations, aiming for a qualitative understanding of the general features of the FO and checking the relation to the former models of this process \cite{old_TL_FO,Mo05a,Mo05b}, while for the more quantitative calculations and comparison with the experimental data we will restrict our modeling to midrapidity region \cite{CsMag_future}.

%%%%%%%%%%%%%%%%%%%%%%%%%%%%%%%%%%%%%%%%%%%%%%%%%%%%%%%%%%%%%%%%%%%%%%%%%%%%%%%%%%%%%%%%%%%%%%%%%%%%%%%%%%%%%%%%%%
\begin{figure*}[htb!]
\centering
\includegraphics[width=0.7\textwidth]{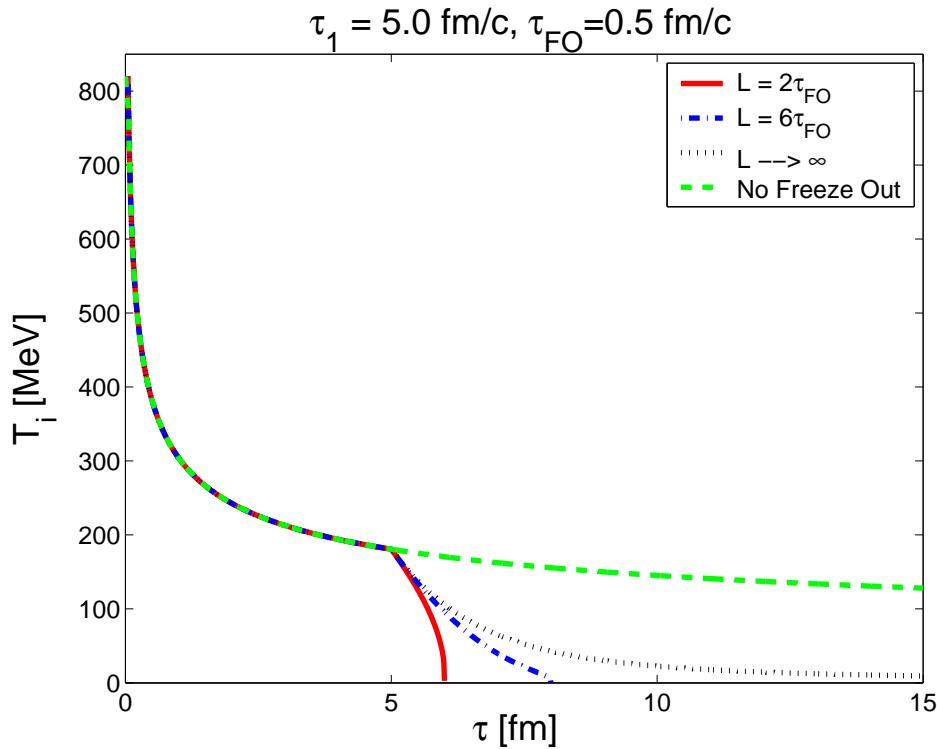}
\caption{Evolution of temperature of the interacting matter for
different FO layers. $T_i(\tau_0=0.05\ fm)=820\ MeV$, $T_{FO}=180\ MeV$. "No Freeze Out" means that we used standard Bjorken hydrodynamics even in phase II.}
\label{fig1}
\end{figure*}
%%%%%%%%%%%%%%%%%%%%%%%%%%%%%%%%%%%%%%%%%%%%%%%%%%%%%%%%%%%%%%%%%%%%%%%%%%%%%%%%%%%%%%%%%%%%%%%%%%%%%%%%%%%%%%%%%%

Thus, finally, we have the following simple model of fireball created in relativistic heavy ion collision.\\
{\bf Initial state, $\tau=\tau_0$}\ \  $e_0,\ n_0$\\
{\bf Phase I, Pure Bjorken hydrodynamics, $\tau_0\le \tau\le \tau_1$}
\beq
e(\tau)=e_0\left(\frac{\tau_0}{\tau}\right)^{1+c_o^2}\,, \quad
n(\tau)=n_0\left(\frac{\tau_0}{\tau}\right)
\eeq{pure_bjor2}
{\bf Phase II, Bjorken expansion and gradual FO,  \\ $\tau_1\le \tau\le \tau_1+L$}\\
Solving Eqs. (\ref{eint},\ref{nint}) we obtain: 
\beq
e^i(\tau)=e_0\left(\frac{\tau_0}{\tau}\right)^{1+c_o^2}\left(\frac{L+\tau_1-\tau}{L}\right)^{L/\tau_{FO}}\,, 
\eeq{bjor_FO_1}
\beq
n^i(\tau)=n_0\left(\frac{\tau_0}{\tau}\right)\left(\frac{L+\tau_1-\tau}{L}\right)^{L/\tau_{FO}}\,.
\eeq{bjor_FO_2}
The difference with respect to the pure Bjorken solution (\ref{pure_bjor2}) is in the last multiplier, and we see that, as expected, the interacting component completely disappears when $\tau$ reaches $\tau= L+\tau_1$.

With these last equations we have completely determined evolution of the interacting component \cite{Bjorken_FO,Bjorken_FO_mass}. Knowing $e^i(\tau)$ and EoS we can find temperature, $T_i(\tau)$. Due to symmetry of the system $u_i^\mu(\tau)=u^\mu(\tau_0)=(1,0,0,0)$. Finally, $f^i(\tau)$ is a thermal distribution with given $T_i(\tau)$, $n^i(\tau)$, $u_i^\mu(\tau)$.

\begin{figure}[htb!]
\centering
\includegraphics[width=0.48\textwidth]{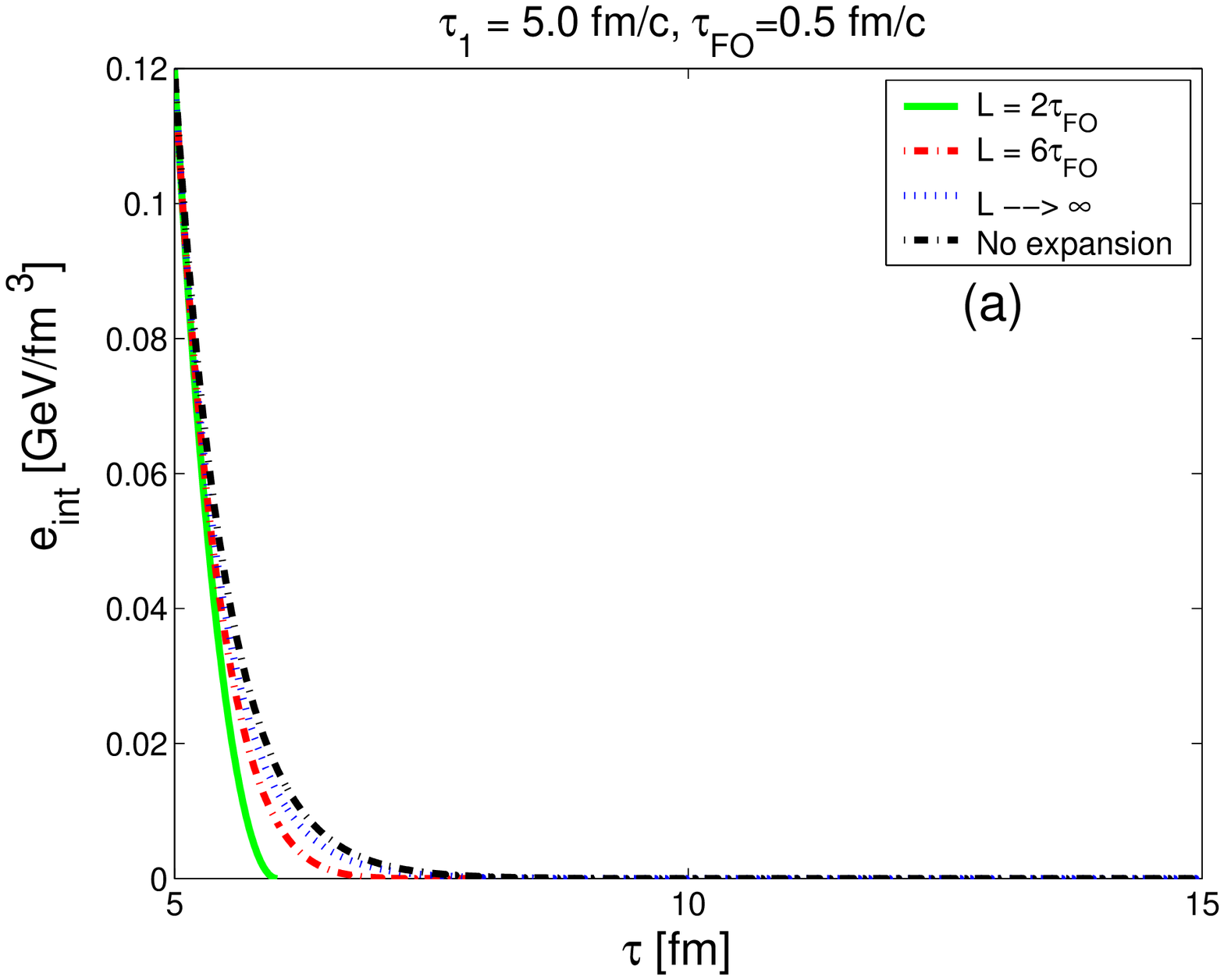} 
\includegraphics[width=0.48\textwidth]{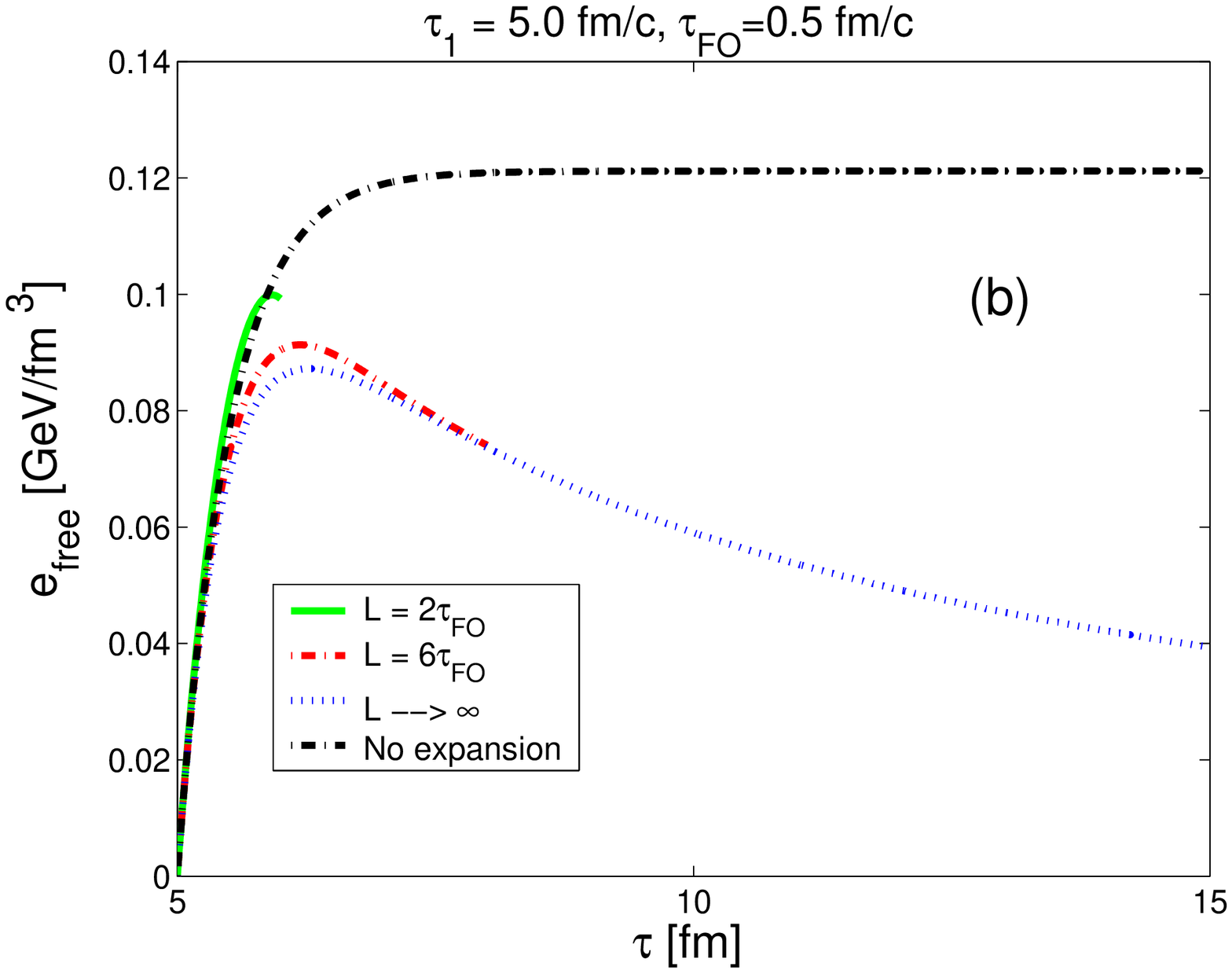}
\caption[]{Evolution of the energy densities for the interacting (a) and free  (b) components for
different FO layers during phase II. $T_i(\tau_0=0.05\ fm)=820\ MeV$, $T_{FO}=180\ MeV$.  }
%\vspace{-0.7cm}
\label{es}
\end{figure}

However, for us the more interesting is the free component, which is the source of the observables. Eqs. (\ref{efree},\ref{nfree}) give us the evolution of the $e_f$ and $n_f$, and one can easily check that these two equations are equivalent to the following equation on the distribution function:
\beq
\frac{d f^f}{d \tau}=-\frac{f^f}{\tau}+\frac{f^i}{\tau_{FO}}\frac{L}{L+\tau_1-\tau}\,.
\eeq{ffree}
The measured post FO spectrum is given by the distribution function at the outer edge of the FO layer, i.e. by $f^f(L+\tau_1)$. 

Most of the observables will depend only on this momentum distribution. However the two particle correlations depend also on the space-time origins of the detected particles. Thus, in order to calculate two particle correlations we have to keep track on both  the momentum and the space-time coordinates of the freeze out point of particles (this can be done, for example, in the source function formalism), i.e. we can use the  full information regarding the space-time evolution of free component, eq. (\ref{ffree}), and correspondingly can get some restrictions on it from the data.    

%%%%%%%%%%%%%%%%%%%%%%%%%%%%%%%%%%%%%%%%%%%%%%%%%%%%%%%%%%%%%%%%%%%%%%%%%%%%%%
\section{Results from the model}
%%%%%%%%%%%%%%%%%%%%%%%%%%%%%%%%%%%%%%%%%%%%%%%%%%%%%%%%%%%%%%%%%%%%%%%%%%%%%%

Aiming for a qualitative illustration of the FO process we show below the results for the ideal massive pion gas with J\"uttner equilibrated distribution \cite{Juttner}: 
\beq
f^i(\tau,\vec{p})=\frac{g}{(2\pi)^3}e^{-p^\mu u_\mu /T_i(\tau)}=\frac{g}{(2\pi)^3}e^{-\sqrt{|\vec{p}|^2+m_\pi^2}/T_i(\tau)}\,,
\eeq{Jut}
in the comoving frame, where $u^\mu=(1,0,0,0)$. Here the degeneracy of pion is $g=3$, while the baryon chemical potential in the case of pions is zero. 
\\ \indent
Contrary to the illustrative example in \cite{Bjorken_FO} here we do not neglect the pion mass. 
During FO the temperature of the interacting component decreases to zero, so at late stages 
of the FO process this new calculation is better justified. 
We will see below that $T_i$ falls below $m_\pi$ quite soon, and so the J\"uttner distribution is a good approximation of the proper Bose pion distribution  \cite{Bjorken_FO_mass}. 
\\ \indent
For our system we have the following EoS:
$$
e^i=\frac{3g}{2\pi^2}m^2T_i^2K_2(a)+\frac{g}{2\pi^2}m^3T_i K_1(a)\,, 
$$
\beq
P^i=\frac{g}{2\pi^2}m^2T_i^2K_2(a)\,,
\eeq{EoS}
where $K_n$ is Bessel function of the second kind, and $a=m/T_i$.

Eqs.  (\ref{eint}) and (\ref{EoS}) result into the following equation for the evolution of the temperature of the interacting component:
$$
\frac{dT_i}{d \tau}= -\frac{T_i}{\tau} \frac{4 T_i^2 K_2(a) + m T_i K_1(a) }
{12 T_i^2 K_2(a) + 5 m T_i K_1(a) + m^2 K_0(a)}
$$
\beq
- \frac{T_i}{\tau_{FO}} \left(\frac{L}{L+\tau_1-\tau} \right)\frac{3 T_i^2 K_2(a) + m T_i K_1(a) }
{12 T_i^2 K_2(a) + 5 m T_i K_1(a) + m^2 K_0(a)}\,.
\eeq{T_eq}
Furthermore, we have used the following values of the parameters: $\eta_R=4.38$, $A_{xy}=\pi R_{Au}^2$, where $ R_{Au}=7.685$ fm is the $Au$ radius, $\tau_0=0.05\ fm$, $T_i(\tau_0)=820\ MeV$, $\tau_1=5\ fm$, what leads to $T_i(\tau_1)=T_{FO}=180\ MeV$, and  $\tau_{FO}=0.5\ fm$. During the pure Bjorken case the evolution of the temperature is govern by eq. (\ref{T_eq}) without the second (freeze out) term on the r.h.s. 

In Fig. \ref{fig1} we present the evolution of the temperature of the interacting matter, $T_i(\tau)$, for different values of FO time $L$, and
Figs. \ref{es} present the evolution of the energy densities for the interacting and free components. These satisfy the energy conservation $E_f(\tau)+E_i(\tau)+W(\tau)= E(\tau_0)$, where $E_f(\tau)$ and $E_i(\tau)$ are given by eq. (\ref{Etot}) with corresponding energy density.  

\begin{figure*}[htb!]
\centering
\includegraphics[width=10.cm, height =8.0cm]{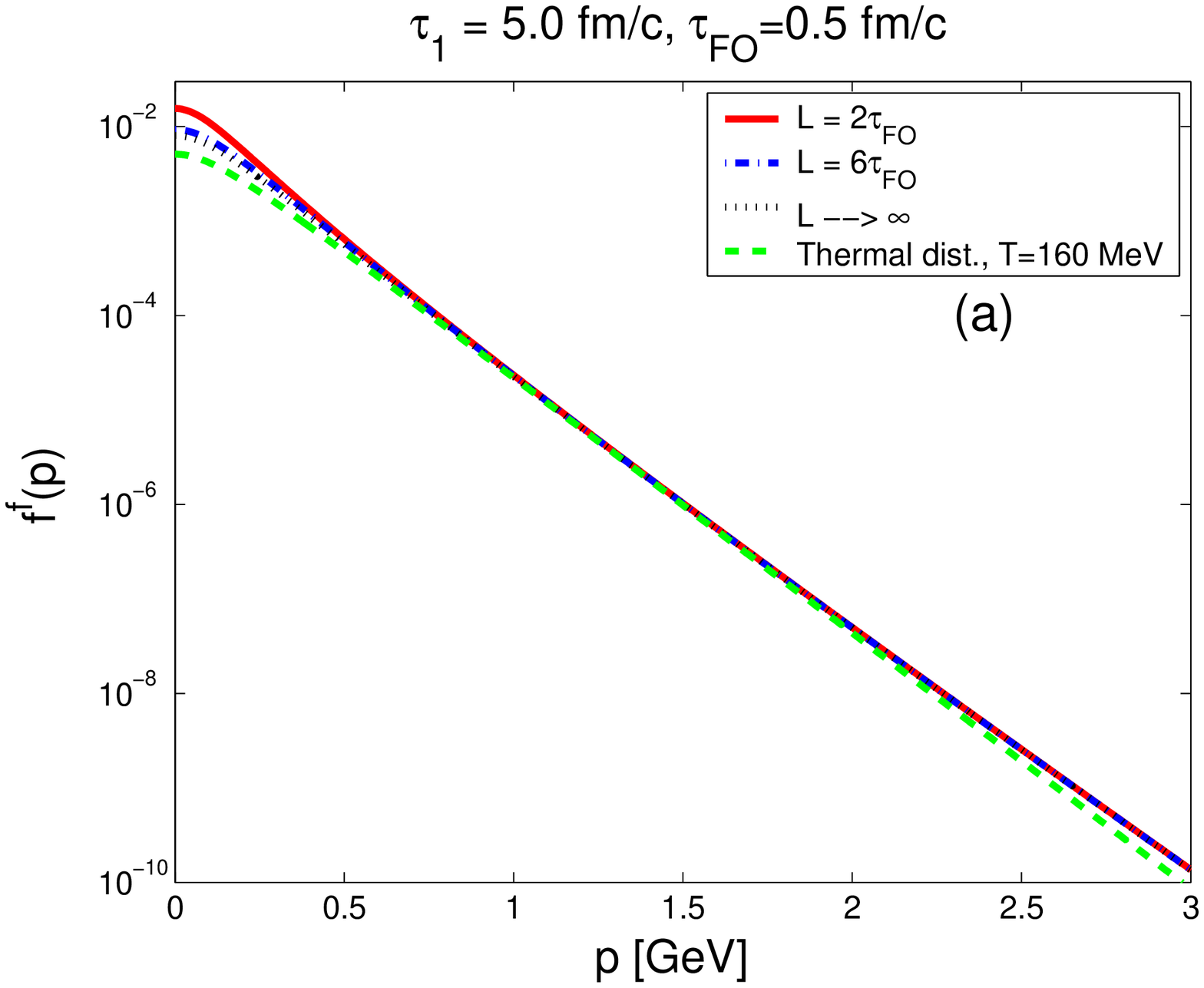} 
\includegraphics[width=6.65cm, height =7.55cm]{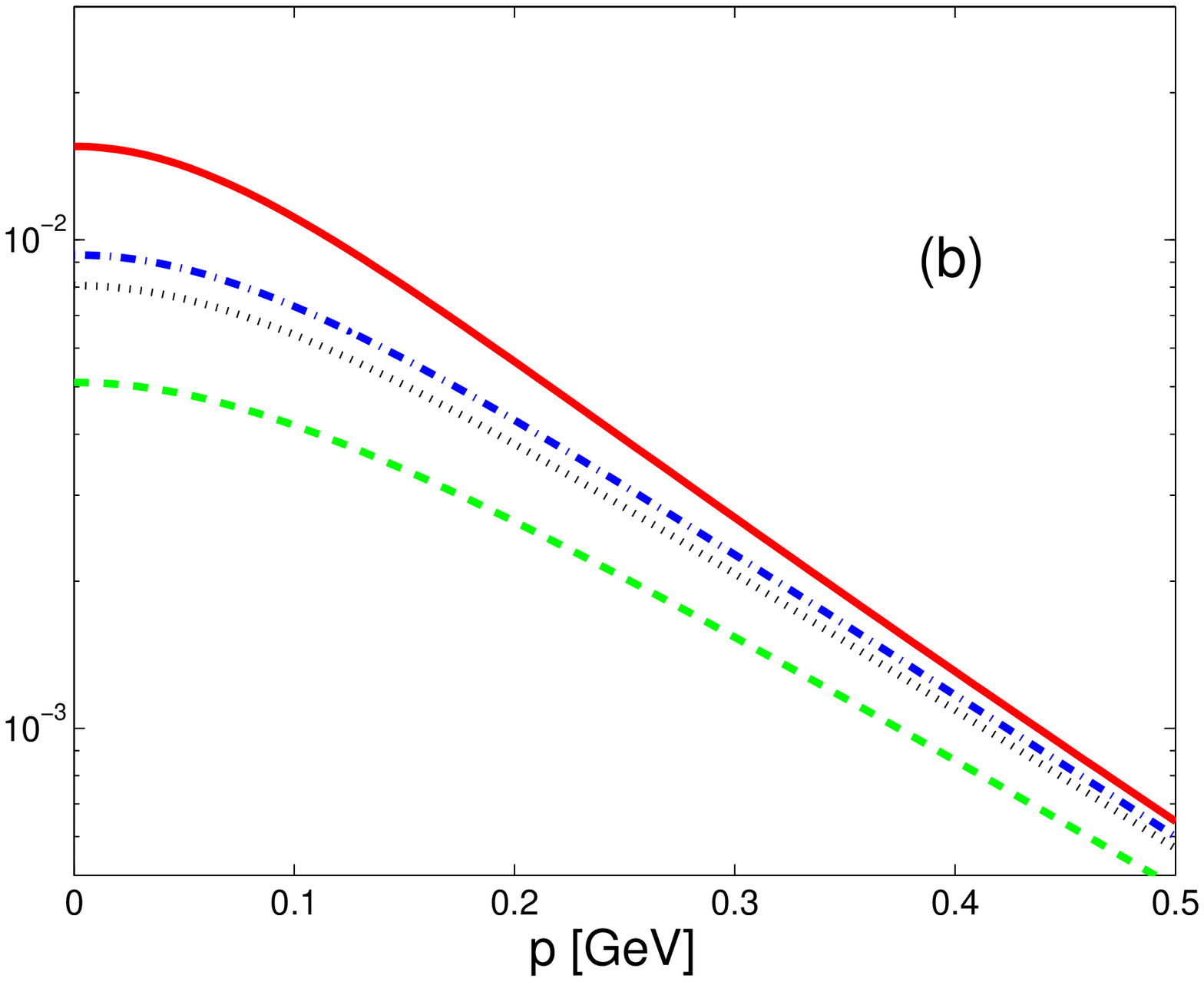}
\caption[]{Final post FO distribution for different FO layers as a function of the momentum in the
FO direction, $p=p^x$ in our case ($p^y=p^z=0$). The initial conditions are specified in the text. Dashed curve shows thermal distribution with temperature $T=160\ MeV$. Subplot (b) is a magnified view on the low momentum region of subplot (a). 
}
\label{fig2}
\end{figure*}

\begin{figure*}[htb!]
\centering
\includegraphics[width=10.8cm, height =8.20cm]{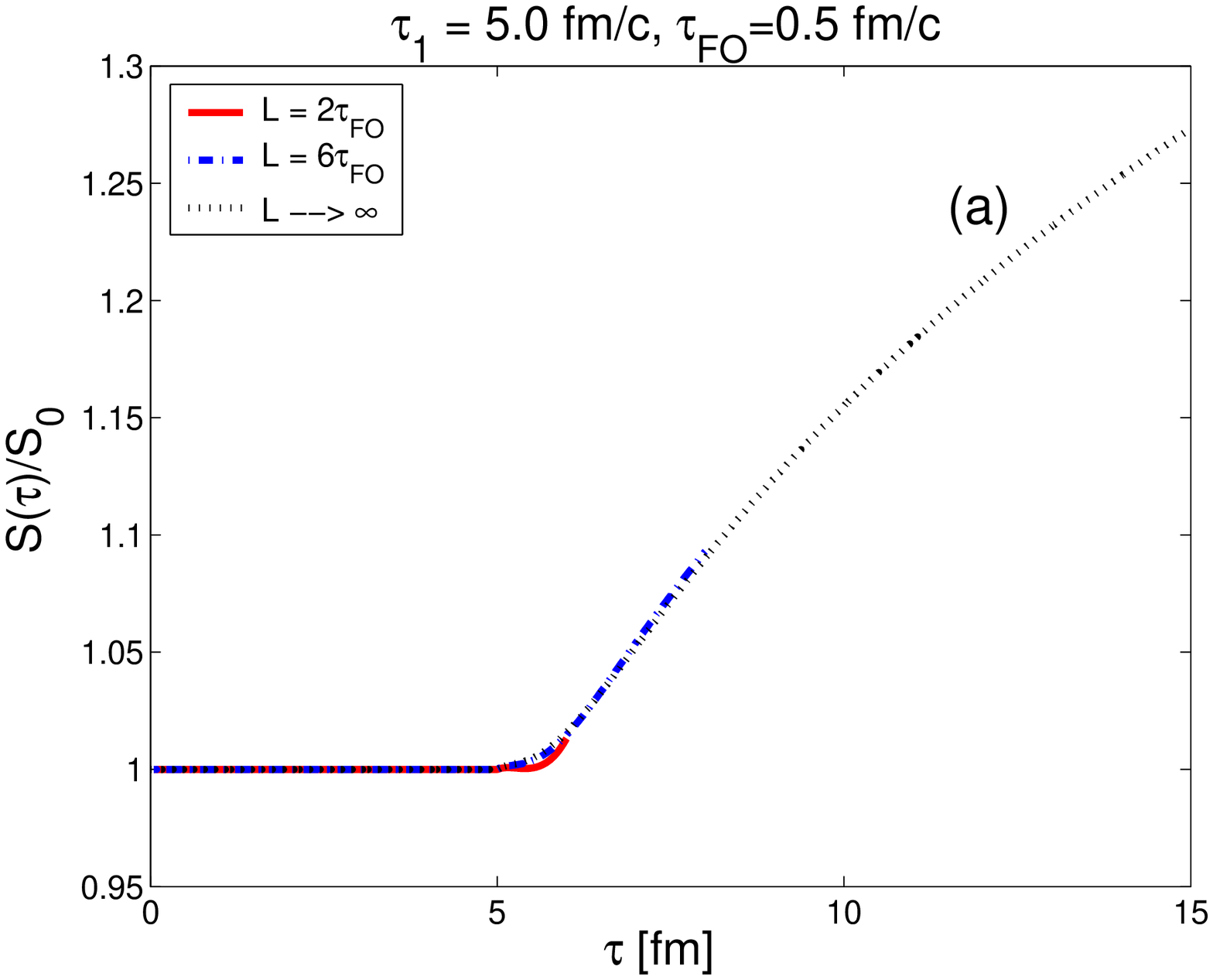} 
\includegraphics[width=6.0cm, height =7.8cm]{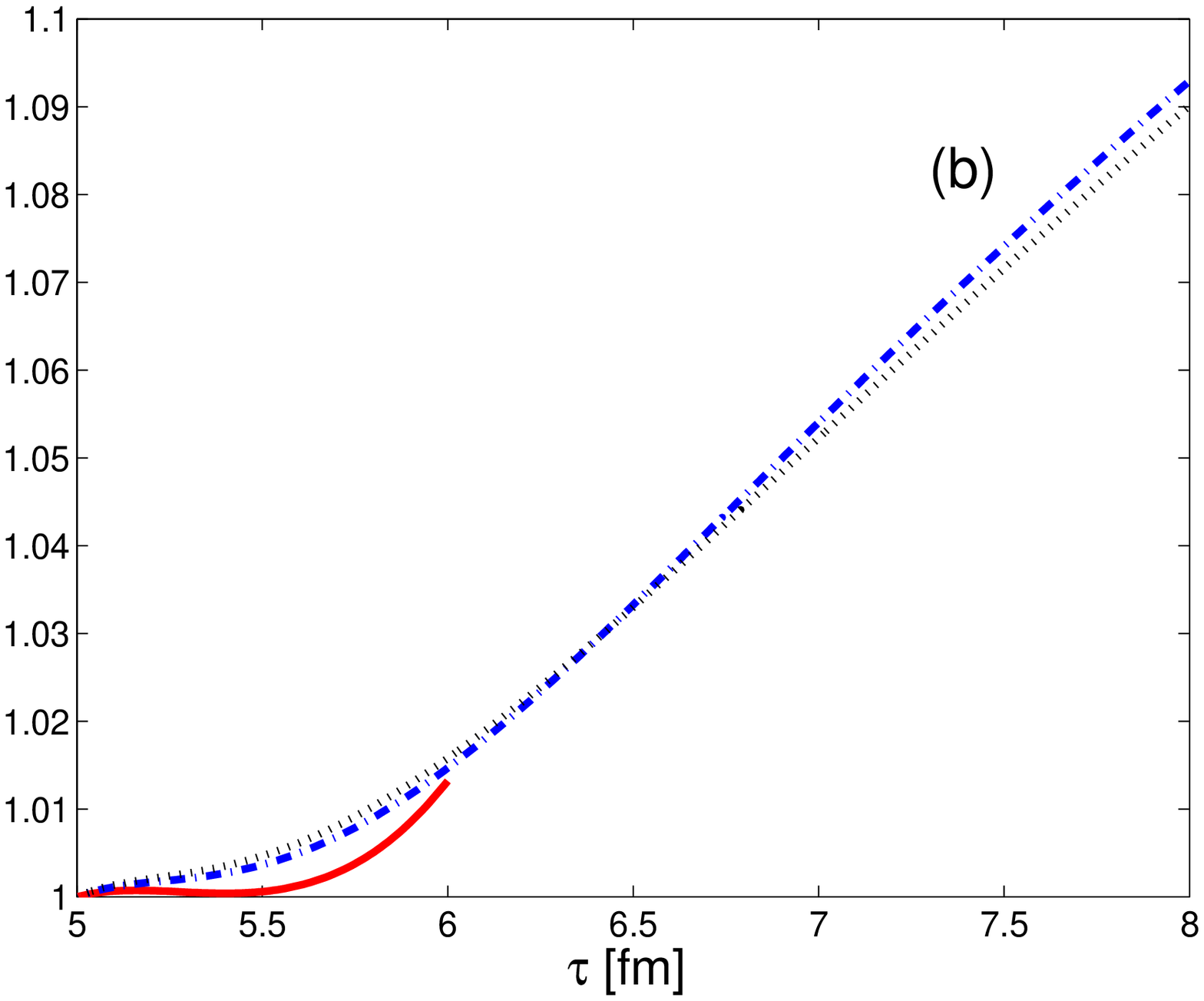}
\caption{Evolution of the total entropy for
different FO layers.  The initial conditions are specified in the text. Subplot (b) is a magnified view on the beginning of FO region of subplot (a). }
\label{entrop}
\end{figure*}

As it was already shown in Ref.\cite{old_TL_FO,Mo05b}, the final post FO particle distributions, shown on Fig. \ref{fig2},
are non-equilibrated distributions, which deviate from thermal ones particularly in the low momentum region.
By introducing  and varying the thickness of the FO layer, $L$, we are strongly affecting the evolution
of the interacting component, see Fig. \ref{fig1}, but the final post FO distribution shows strong universality:
for the FO layers with a thickness of several $\tau_{FO}$, the post FO distribution already looks very close to that for an infinitely long FO calculations, see Fig. \ref{fig2} left plot. Differences can be observed only for the very small momenta, as shown in  Fig. \ref{fig2} right plot.
So, the inclusion of the expansion into our consideration does not smear out this very important feature of the 
gradual FO.
\\ \indent
Please note, that if one would look on our post FO distributions only in the medium momenta regions (with $0.5\ GeV<|p|< 2.5\ GeV$), then one could fit these spectra 
reasonable well with equilibrated distribution with temperature $T=160\ MeV$, see dashed line on Fig. \ref{fig2}. However, for low and high momenta such a fit would strongly disagree, this has been shown already in earlier FO model in Ref. [7b].% \cite{Mo05a}.
\\ \indent
For our illustrative calculation we assumed $\eta_R=4.38$ what is one unit smaller than the original rapidity of the colliding nuclei for $100\ GeV*A$. The experimental data show a smaller plato in rapidity, but we used a classic one-dimensional Bjorken model without transverse expansion which correspondingly requires a larger longitudinal size. 
\\ \indent
It is important to check the non-decreasing entropy condition \cite{Bjorken_FO,Bjorken_FO_mass,cikk_2} to see whether such 
a process is physically possible.
Figs. \ref{entrop} present the evolution of the total entropy, $S(\tau)$, calculated based on the full 
distribution function, $f(\vec{p})=f^i(\vec{p})+f^f(\vec{p})$:
\beq
s(\tau)=\int d^3 p f(\tau) \left[ 1 - \ln \left( \frac{(2\pi)^3}{g} f(\tau) \right)\right]\,. 
\eeq{s_den}
The total entropy is then given by eq. (\ref{Stot}).

\indent
From this figure we can make an important conclusion, that gradual freeze out with rethermalization produces entropy. This is not new - in Ref. \cite{grassi04} the authors
reached the same conclusion based on longitudially boost invariant hydrodynamical model with continuous emission (what means infinitely long FO in our language) with different escape probability \cite{vol_em}. In Ref.  \cite{vol_em,grassi04} $P_{esc}$ is given by the probability to escape without further collisions, 
calculated based on Glauber formula.
In the case of complete rethermalization of the interacting component, which we also assume, their entropy increase can go up to $70\% $! This leads then to the corresponding increase in the pion production, which is a rough measure of entropy. Certanly such a hydro+continuous emission model leads the nonthermal final distributions, and authors found a reasonable agreement with NA35 transverse mass spectra. This model naturally explains the large number of the produced pions, which was a trouble for models with sharp FO hypersurface.      

\indent
In our model, as well as in \cite{vol_em,grassi04}, there are two main sources of entropy production, namely the separation process (into interacting or free component) and the rethermalization process. Separation process leads to entropy increase because particles have now more accessible states being either interacting or free, not just interacting. The entropy increase in the rethermalization process is a manifistation of the H theorem.

\indent
At the late stages of the long FO this entropy increase can be approximated analytically in our model.
$$
\frac{ds(\tau)}{d \tau}=\int d^3 p \frac{df(\tau)}{d\tau} \left[ 1 - \ln \left( \frac{(2\pi)^3}{g} f(\tau) \right)\right] -
$$
\beq
 - \int d^3 p  \frac{df(\tau)}{d\tau} \,. 
\eeq{ds}

\indent
At the late stages of the FO, when $\tau$ is close to the end of FO $L+\tau_1$, for long FO, i.e. $L>>\tau_{FO}$, the interacting component, $f_i$, can be neglected next to free component, $f_f$, (see eqs. (\ref{bjor_FO_1},\ref{bjor_FO_2}) and Fig. \ref{es} ), and so, using eq. (\ref{ffree}) we obtain 
\beq
\frac{d f}{d \tau}\approx \frac{d f^f}{d \tau} \approx-\frac{f^f}{\tau}\,.
\eeq{fff} 
Then, putting eq. (\ref{fff}) into eq. (\ref{ds})  we get the following equation for the entropy evolution:
\beq
\frac{ds(\tau)}{d \tau}\approx -\frac{s(\tau)}{\tau}  +  \frac{1}{\tau} \int d^3 p  f^f \,.
\eeq{ds2}
If there is a baryon charge (or other conserved charges) in the system in general case the last term in the eq. (\ref{ds2}) can then  be related to $n_f$: 
\beq
\frac{ds(\tau)}{d \tau}\approx -\frac{s(\tau)}{\tau}  + \frac{n^f(\tau)}{\tau}\,,
\eeq{ds2b}
and for the baryon density, eq. (\ref{nfree}), in this limit we have
\beq
\frac{dn^f(\tau)}{d \tau}\approx -\frac{n^f(\tau)}{\tau}\quad \Rightarrow \quad n^f(\tau)=n_{FO} \frac{L+\tau_1}{\tau} \,,
\eeq{dn2}
where $n_{FO}$ is the free baryon density at the end of FO. 

So, for the total entropy, eq. (\ref{Stot}), we obtain:
\beq
\frac{dS}{d \tau}=2 A_{xy} \eta_{R} \frac{d\left(s  \tau \right)}{d \tau } \approx -\frac{S}{\tau}  + \frac{N_f}{\tau} + \frac{S}{\tau} = \frac{ N_{FO} }{\tau} \,,
\eeq{ds_fin}
where $N_f(\tau)$ is the total number of frozen out baryon charges, which is in our limit, eq. (\ref{dn2}), equal to the total number of frozen out baryon charges  at the end of FO, $N_{FO}$, which is, due to conservation law, the same as initial number of interacting baryon charges.
\beq
S(\tau)=S_{FO} + N_{FO} \ln \frac{\tau}{L+\tau_1} \,,
\eeq{S_fin}
where $S_{FO}$ is the final entropy. 
\\ \indent
Thus, we see that the total entropy increases during simultaneous expansion and gradual FO, as it is clearly seen on Fig. \ref{entrop} for the $L\rightarrow\infty$ curve. If there is a baryon charge in the system, then at the very
end of the FO process entropy increases logarithmically, according to eq. (\ref{S_fin}).
\\ \indent
On the other hand the fast FO in the narrow layer, which can be studied in our model, contrary to \cite{vol_em,grassi04,vol_em_2}, does not lead to entropy increase. Already for the FO layer as thin as $2\tau_{FO}$ the entropy production is negligible, around $1\% $, see Fig. \ref{entrop}.

%%%%%%%%%%%%%%%%%%%%%%%%%%%%%%%%%%%%%%%%%%%%%%%%%%%%%%%%%%%%%%%%%%%%%%%%%%%%%%
\section{Conclusions}
%%%%%%%%%%%%%%%%%%%%%%%%%%%%%%%%%%%%%%%%%%%%%%%%%%%%%%%%%%%%%%%%%%%%%%%%%%%%%%
In this paper we presented a model, which  describes simultaneously freeze out and Bjorken expansion, and thus, it is more a physical extension of the oversimplified FO models without expansion \cite{old_SL_FO_2,old_SL_FO,old_TL_FO,Mo05a,Mo05b}, which allowed us 
to study FO in a layer of any thickness, $L$, from $0$ to $\infty$.
Another important feature of the proposed model is that it connects the pre FO hydrodynamical quantities, 
like energy density, $e$, and baryon density, $n$, with the post FO distribution function in a relatively simple way. As a zero order approach to the real physical system we studied the FO of the  
pion gas \cite{Bjorken_FO_mass}.
\\ \indent
The results show that the inclusion of the expansion into FO model, although strongly affects the evolution
of the interacting component, does not smear out the universality of the final post FO distribution, observed already in Refs. \cite{old_TL_FO,Mo05a,Mo05b}.
\\ \indent
Another important conclusion of this work is stressing once again the importance to always check the 
non-decreasing entropy condition  \cite{Bjorken_FO,Bjorken_FO_mass,cikk_2}, since long gradual freeze 
out may produce substantial amount of entropy, as shown in Fig. \ref{entrop}.

This may have important consequence for QGP search. For many qualitative estimations it was assumed that all the entropy is produced at the early stages of the reaction and that   the expansion, hadronization and FO go adiabatically, and thus number of pions can serve as a rough measure of entropy. However, if a non-negligible part of the entropy, say $10 \%$, is produced during FO, then some estimations, for example of strangeness vs entropy (pion) production %\cite{Gorenstein,Gadzicki}
have to be reviewed.

%%%%%%%%%%%%%%%%%%%%%%%%%%%%%%%%%%%%%%%%%%%%%%%%%%%%%%%%%%%%%%%%%%%%%%%%%%%%%%
\section{Aknowledgements}
%%%%%%%%%%%%%%%%%%%%%%%%%%%%%%%%%%%%%%%%%%%%%%%%%%%%%%%%%%%%%%%%%%%%%%%%%%%%%%
Authors thank E. Molnar for fruitful discussions. This work was partially
supported by Grant No. FIS2005-03142 from MEC (Spain) and FEDER.

\end{document}